\documentclass[12pt,twoside,dvips]{article}
\usepackage{verbatim}
\usepackage{amssymb}
\usepackage{amsfonts}
\usepackage{amsmath}
\usepackage{pifont}
\usepackage[spanish,english]{babel}
\usepackage[pdftex]{graphicx}
\usepackage{verbatim}
\usepackage{epsfig}
\usepackage{bm}
\usepackage{longtable}

\begin{document}

\title{Seesaw mechanism with Yukawa alignment for neutrinos}
\author{R. Mart\'{\i}nez$\thanks{%
e-mail: remartinezm@unal.edu.co}$, F. Ochoa$\thanks{%
e-mail: faochoap@unal.edu.co}$,  M. O. Ospina \and Departamento de F\'{\i}sica, Universidad
Nacional de Colombia, \\ Ciudad Universitaria, Bogot\'{a} D.C.}
 
\maketitle

\begin{abstract}
In the extension of the standard model with one right-handed neutrino and one Higgs triplet, we propose a suppression mechanism, obtaining small masses for the active neutrinos, while mixing angles are predicted with a right-handed neutrino at the TeV scale and Yukawa couplings at the order of $\mathcal{O}(1)$. In this extension, the seesaw formula is proportional to the difference between two Yukawa couplings: the one that governs the interactions of the ordinary matter through the Higgs triplet, and the coupling of the new neutrino through the scalar doublet, so that by aligning both Yukawa couplings, exact zero-mass active neutrinos are obtained. By perturbating this alignment condition, we obtain neutrino masses proportional to the magnitude and direction of the perturbation in the flavour space. Bimaximal and nearly bimaximal mass structures emerge from specific unalignment forms.
\end{abstract}

\section{Introduction}

Although there are not direct measurements of the absolute values of neutrino masses, we have compelling evidences that they are different from zero, which force us some type of extension of the standard model (SM). In particular, the effect of flavour oscillation observed in solar, atmospheric and reactor neutrinos \cite{solar, atmospheric, reactor, oscil-other}, provide us a set of data related to the neutrino masses and flavour mixing. As a result of these experiments and different global fits \cite{fits}, we obtain the following general features:

\begin{itemize}
\item The three active neutrinos must be light, with masses at the orders of eV - KeV. 
\item The data do not distinguish among three mass patterns for three massive neutrinos $\nu _1$, $\nu _2$ and $\nu _3$: the normal hierarchy (NH), where $m_1 \ll m_2 \ll m_3$, the inverted hierarchy (IH) with $m_3 \ll m_1 \lesssim m_2$, and the quasidegenerate pattern (QD) where $m_1 \approx m_2 \approx m_3$.
\item The data show preference for three nonzero mixing angles, where two of them are large, while the third is small.
\item There are not still convincing evidences for CP violation in the leptonic sector, thus a CP complex phase in the mixing matrix is unknown.
\end{itemize}

For the first feature, the neutrinos may become massive through the spontaneous breaking of the gauge symmetry if  the particle spectrum of the SM is extended. In many of these extensions, the smallness of the masses can be understood in the framework of a seesaw mechanism, where the extra particle content induces a suppression factor in the mass terms of the neutrinos. For example, in the most simple see-saw mechanism (type I) \cite{seesaw1}, the introduction of a right-handed sterile neutrino with mass $M_R$ leads us to a light neutrino mass of the form $M_{\nu }=f^2\upsilon ^2/M_R$, where $f$ is the Yukawa coupling of the new neutrinos with the SM leptons, which usually is assumed of the order unity, and $\upsilon =246$ GeV the electroweak vacuum expectation value (VEV). In order to obtain massive neutrinos at the eV scale, we must have heavy right-handed neutrinos at the order of $M_R\sim 10^{14}$ GeV, which exclude any possibility to be produced in current or future accelerators. Another alternative (see-saw type II) is by introducing a Higgs triplet with VEV $\upsilon _{\Delta}$ \cite{seesaw2}, which couple to the SM left-handed leptons through the Yukawa coupling $\Gamma$. In this case, the seesaw formula has the form $M_{\nu }=\Gamma \upsilon _{\Delta}$, generating mass values at the order of the VEV of the Higgs triplets, which can be small. As a consequence to add this higher-dimensional scalar sector, the SM $\rho _0$ parameter is modified, which roughly imposes an upper limit of the order of

\begin{eqnarray}
\upsilon _{\Delta} \lesssim 7 \text{ GeV}.
\label{rho-constraint}
\end{eqnarray}
Although there are not any lower limit for this parameter, the masses of the extra charged Higgs bosons predicted by the model depends as $\upsilon _{\Delta} ^{-1}$. Thus, a VEV as small as the eV scale may generate too heavy extra Higgs particles, unattainable to experimental verification. The seesaw mechanism type III \cite{seesaw2} produces the same neutrinos mass as in the type I, but with a right-handed lepton triplet. Finally, in the so called inverse see-saw mechanism \cite{seesawinverse}, the addition of a vector-like sterile singlet neutrino gives masses of the form $M_{\nu }=\mu f^2\upsilon ^2/M^2$, where $\mu $ is a Majorana mass term induced by the left-handed component of the lepton singlet and $M$ its Dirac mass. In this case, the neutrino masses have an additional suppresion factor if $\mu$ is small, obtaining light neutrinos at the eV scale for heavy neutrinos at the TeV scales.

Regarding the large mixing structure, this can be obtained if the original mass matrix of the light neutrinos exhibits specific structures in their components. Globally, the data from solar and atmospheric oscillations favour a bimaximal and tribimaximal mixing matrices \cite{barger1}. Assuming mixing only in the neutral sector, the mass matrix that gives a bimaximal mixing have the general form \cite{barger}:

\begin{eqnarray}
M_{\nu \text{b} }=m_0\left[\begin{pmatrix}
0 & 0  & 0 \\
0 & 1  & -1 \\
0 & -1  & 1 
\end{pmatrix}+
\begin{pmatrix}
2\alpha & \beta & \beta \\
\beta & \alpha  & \alpha \\
\beta & \alpha  & \alpha 
\end{pmatrix}\right],
\label{bimax}
\end{eqnarray}
%
where the coefficients are: 

\begin{eqnarray}
m_0&=&\frac{m_3}{2}, \ \ \ \ \alpha = \frac{m_1+m_2}{4m}, \ \ \ \ \beta=\sqrt{2}\left(\frac{m_1-m_2}{4m}\right), \ \ \ 
\label{coeff}
\end{eqnarray}
In particular,  $|\beta|, \alpha \ll 1$ for a NH scheme,  $\alpha \gg 1$ for the IH, and $|\beta |\ll \alpha \approx 1$ for  the QD pattern. There are other forms that can be obtained by perturbating the above bi- and tribi-maximal matrices \cite{perturbation}..

In this paper, we consider a SM extension with one Higgs triplet and one right-handed neutrino, which lead us to a combined type I and type II seesaw mechanism for the light neutrino masses. However, we additionally propose a variation of these mechanisms, where the soft-breaking of an alignment condition between the new Yukawa couplings produces an additional suppression factor, predicting masses at the eV scale for the active neutrinos, while the sterile neutrino remains at the TeV scale, the VEV of the Higgs triplet at the GeV scale and the new Yukawa couplings at $\mathcal{O}(1)$. Also, by assuming deviations to the alignment condition, a mass matrix nearly bi-maximal is obtained naturally, and mixing angles and mass differences can be fitted with few free parameters.

\section{The alignment condition\label{sec:alignment}}

\subsection{Survey of the model}

The model corresponds to the usual SM spectrum with the addition of one right-handed neutrino $N_R$ and one Higgs triplet $\Delta$, which has VEV $\langle \Delta \rangle=\upsilon _{\Delta}$ and hipercharge 2. These new particles feature the following $(SU(3)_c,SU(2)_L, U(1)_Y)$ representations:

\begin{eqnarray}
N_R: \left(1,1,0\right), \ \ \ \ \  \ \ \  \ \ \ \ \ \ \ \ \ \Delta =
\begin{pmatrix}
\frac{1}{\sqrt{2}}\upsilon _{\Delta}+\Delta ^{0} & \frac{1}{\sqrt{2}}\Delta ^{+}\\
\frac{1}{\sqrt{2}}\Delta ^+  & \Delta ^{++}
\end{pmatrix}:\left(1,3, 2\right).
\label{newpart}
\end{eqnarray}

The most general renormalizable Yukawa Lagrangian compatible with the $SU(2)_L \times U(1)_Y$ symmetry is:

\begin{eqnarray}
-\mathcal{L}_Y&=& -\mathcal{L}_{Y(SM)}+f_{i}\overline{\ell^{i}_{L}} \widetilde{\phi} {N}_{R}+  \frac{1}{2}M_R \overline{N_{R}^c} N_{R}+\frac{1}{2}\Gamma_{ij} \overline{\ell^{ic}_{L}}\Delta \ell^{j}_{L} + \text{H.c}, 
\label{yukawa}
\end{eqnarray}  
where a sum over repeated flavour indices ($i,j = 1, 2, 3$) is implied, $\phi $ is the SM Higgs doublet with VEV $\upsilon = 246$ GeV, $\widetilde{\phi}=i\tau _2 \phi ^*$ and $\Gamma _{ij}$ is symmetric due to the Fermi statistics. In particular, after the symmetry breaking, we obtain the mass terms for the neutral sector, which can be written as:


 \begin{eqnarray}
 -\langle \mathcal{L}_0 \rangle&=&\overline{\chi _L}M\chi _R + \text{H.c} \nonumber \\
&=& \frac{1}{2\sqrt{2}}\left(\overline{\nu ^{i}_{L}}, \overline{N_{R}^c} \right)
 \begin{pmatrix}
\upsilon _{\Delta}\Gamma _{ij} & \upsilon f_i \\ \\
\upsilon f_j  & \sqrt{2}M_R 
 \end{pmatrix}
\begin{pmatrix}
\nu ^{jc}_{L}  \\
N_R
\end{pmatrix}+ \text{H.c}.
\label{massterms2}
\end{eqnarray}   
where $\Gamma _{ij}$ label a $3\times 3$ matrix, and $f_i$ a three dimensional vector. Since the matrix $M$ can be complex, it is convenient to calculate the squared mass matrix in order to guarantee real masses, obtaining:

\begin{eqnarray}
M^2=MM^{\dag}=\frac{1}{8}\begin{pmatrix}
\upsilon _{\Delta}^2\Gamma _{ik}\Gamma _{jk}^{\ast}+\upsilon ^2f_if_j^{\ast} & \upsilon _{\Delta}\upsilon \Gamma _{ik}f_k^{\ast}+\sqrt{2}\upsilon f_iM_R^{\ast} \\
 \upsilon _{\Delta}\upsilon f_k\Gamma _{jk}^{\ast}+\sqrt{2}\upsilon M_Rf_j^{\ast} & \upsilon ^2 \left|f\right|^2+2M_R ^2 
 \end{pmatrix},
 \label{mass-matrix}
\end{eqnarray}
where a sum over the index $k=1,2,3$ is assumed and $|f|^2=\sum _{i}|f_i|^2$ is the magnitude of the $\nu_L$-$N_R$ coupling.

\subsection{Yukawa coupling alignment}

Although there is not any initial constraint that relates both VEVs $\upsilon $ and $\upsilon _{\Delta}$, we can make a feasible guess, where the scale of $\upsilon _{\Delta}$ is estimated as the electroweak scale suppresed by $M_R$ as $\left(\upsilon _{\Delta}/\sqrt{2}\right)\sim \left(\upsilon /\sqrt{2}\right) ^2/M_R$. By introducing a control parameter $\mu$, we obtain:

\begin{eqnarray}
\upsilon _{\Delta}=\mu \frac{\upsilon ^2}{\sqrt{2}M_R},
\end{eqnarray}
which, for $M_R \sim 1$ TeV and $\upsilon _{\Delta} \lesssim  7$ GeV according to (\ref{rho-constraint}), gives us $\mu \lesssim  0.16$. With the above assumption, and defining the suppression factor

\begin{eqnarray}
\epsilon = \frac{\upsilon}{M_R},
\label{eps}
\end{eqnarray}
the matrix (\ref{mass-matrix}) become:

\begin{eqnarray}
M^2&=&\frac{\upsilon ^2}{16\epsilon ^2}\begin{pmatrix}
 A_{ij} & B_{i} \\ \\
 B_j^{\ast} & C
 \end{pmatrix} \nonumber  \\ \nonumber \\
&=& \frac{\upsilon ^2}{16\epsilon ^2}\begin{pmatrix}
\left(2f_if_j^{\ast}+\mu ^2\Gamma _{ik}\Gamma _{jk}^{\ast}\epsilon ^2\right)\epsilon ^2 & \sqrt{2}\left(2f_i+\mu \Gamma _{ik}f_k^{\ast}\epsilon ^2\right)\epsilon \\ \\
 \sqrt{2}\left(2f_j^{\ast}+\mu f_k\Gamma _{jk}^{\ast} \epsilon ^2 \right)\epsilon  & 2\left(2+ \left|f\right|^2\epsilon ^2\right) 
 \end{pmatrix}.
 \label{mass-matrix-2}
\end{eqnarray}
The determinant of the above matrix can be block calculated as:

\begin{eqnarray}
\det\left[M^2\right]&=&\frac{\upsilon ^2}{16\epsilon ^2}C\det\left[A_{ij}-B_iB_j^{\ast}C^{-1}\right] \nonumber \\
&=&\frac{1}{8}\upsilon ^2\epsilon ^2 \det\left[\left|\mu \Gamma -ff\right|^2_{ij}+\frac{1}{2}\left(\left|f\right|^{2}\left|\mu \Gamma \right|^{2}_{ij}-\left|\mu \Gamma f\right|^2 _{ij}\right)\right] ,
\label{deteminant}
\end{eqnarray}
where, for short,  we have used the following notation for the scalar product of the tensors:

\begin{eqnarray}
\left|\mu \Gamma -ff\right|^2_{ij}&=& \left(\mu \Gamma _{ik} - f_if_k\right)\left(\mu \Gamma _{kj} - f_kf_j\right)^{\ast}, \nonumber \\
\left|\mu \Gamma \right|^{2}_{ij}&=&\mu ^2 \Gamma _{ik}\Gamma _{kj}^{\ast}, \nonumber \\
\left|\mu \Gamma f\right|^2 _{ij}&=&\mu ^2 \Gamma _{ik}f_k^{\ast}f_l\Gamma _{lj}^{\ast}.
\label{tensor-magnitude}
\end{eqnarray}

Since the determinant of a matrix is proportional to the product of its eigenvalues, we will obtain at least one zero mass neutrino if the above determinant equals to zero. In particular, we observe that if both Yukawa couplings $\Gamma$ and $f$ align as:

\begin{eqnarray}
\mu \Gamma _{ij}=f_if_j,
\label{alighment-limit}
\end{eqnarray}     
then the tensor products in (\ref{tensor-magnitude}) lead us to:

\begin{eqnarray}
\left|\mu \Gamma -ff\right|^2_{ij}&=&0, \nonumber \\
\left|f\right|^{2}\left|\mu \Gamma \right|^{2}_{ij}-\left|\mu \Gamma f\right|^2 _{ij}&=&0
\label{tensor-magnitude-aligned}
\end{eqnarray}
which nullify the determinant in (\ref{deteminant}), while the zero eigenvalue has multiplicity three. Thus, in the exact alignment limit as defined as in (\ref{alighment-limit}), the three light neutrinos become massless. This suggests us that in addition to the traditional see saw suppression mechanism, a small deviation from the alignment condition of the Yukawa couplings may help to understand the smallness of the light neutrinos, as discussed in the next section below.

\section{Breaking of the alignment condition}

\subsection{Perturbative block diagonalization} 

Before demanding the alignment condition, we must obtain analytical expressions for the light neutrino (active neutrinos) masses by using the expansion method from reference \cite{grimus}. For that, we first observe that each block of the original mass matrix (\ref{mass-matrix-2}) can be separated as powers of $\epsilon$ as:

\begin{eqnarray}
(A)_{ij}&=& (A_2)_{ij}+(A_4)_{ij}=\left(2f_if_j^{\ast}\right)\epsilon ^2+\left(\mu ^2\Gamma _{ik}\Gamma _{jk}^{\ast}\right)\epsilon ^4, \nonumber \\
(B)_i&=&(B_1)_i+(B_3)_i=\left(2\sqrt{2}f_i\right)\epsilon+\left(\sqrt{2}\mu \Gamma _{ik}f_k^{\ast}\right)\epsilon ^3, \nonumber \\
C&=& C_0+C_2=4+\left(2\left|f\right|^2\right)\epsilon ^2.
 \label{mass-matrix-3}
\end{eqnarray}
%
where, for the dominant parts, the block components exhibit the following hierarchical structure due to the small factor $\epsilon $:

\begin{eqnarray}
(A_2)_{ij} \ll (B_1)_i \ll C_0.
\label{hierarchy}
\end{eqnarray}
Thus, we can block diagonalize the mass matrix recursively as shown in appendix \ref{app:block} into one light $3\times 3$ matrix and a heavy mass associated to the sterile neutrino as:

\begin{eqnarray}
V^TM^2V=
\begin{pmatrix}
m^2_{\nu} & 0 \\
0 & m^2_N
\end{pmatrix}.
\end{eqnarray}
Up to order $\epsilon ^4$, the active neutrino sub-matrix $m^2_{\nu}$ expands as:

\begin{eqnarray}
m_{\nu}^2\approx m_{\nu}^2(\epsilon ^2)+m_{\nu}^2(\epsilon ^4),
\label{neutrino-mass-matrix}
\end{eqnarray}
where

\begin{eqnarray}
m_{\nu}^2(\epsilon ^2)=\frac{\upsilon ^2}{16\epsilon ^2}a(\epsilon^{4}), \ \ \ \ \ m_{\nu}^2(\epsilon ^4)=\frac{\upsilon ^2}{16\epsilon ^2}a(\epsilon ^{6}),
\end{eqnarray}
with $a(\epsilon^{4})$ and $a(\epsilon^{6})$ given by equations (\ref{light-mass-4}) and (\ref{light-mass-6}), respectively. After replacing each block from equation (\ref{mass-matrix-3}), we obtain the following components of the active neutrino matrix:

\begin{eqnarray}
\left(m_{\nu}^2(\epsilon ^2)\right)_{ij}&=&\frac{\upsilon ^2}{16}\left|\mu \Gamma - ff \right|^2_{ij}\epsilon ^2, \nonumber \\
\left(m_{\nu}^2(\epsilon ^4)\right)_{ij}&=&-\frac{\upsilon ^2}{64}\left(\left|\mu \Gamma - ff \right|^2_{il}f_lf_j^{\ast }+2\left(\mu \Gamma-ff\right)_{ik}f_k^{\ast}f_l\left(\mu \Gamma -ff\right)_{lj}^{\ast}\right. \nonumber \\
&&\left.+f_if_k^{\ast}\left|\mu \Gamma - ff\right|_{kj}^2 \right)\epsilon ^4.
\label{eps2-4order}
\end{eqnarray}
where we have applied the short notation from equation (\ref{tensor-magnitude}), and a sum over the repeated indices $l$ and $k=1,2,3$ is also assumed.

\subsection{Perturbation of the alignment condition}

By enforcing the exact alignment condition (\ref{alighment-limit}), both contributions in (\ref{eps2-4order}) cancel out  independently, and the active neutrinos become massless, property that extends to all order of $\epsilon $, according to the discussion of the matrix determinant in section \ref{sec:alignment}. Thus, to obtain massive neutrinos, we must produce perturbations induced by small deviations from the alignment of the Yukawa couplings, which can be parametrized through a small complex parameter $\varepsilon _{ij}$ as:

\begin{eqnarray}
\mu \Gamma _{ij}=\left(1+\varepsilon _{(ij)}\right)f_if_j,
\label{alighment-perturb}
\end{eqnarray} 
for each $ij$ component, where the indices into the brackets, $(ij)$, indicates that no summation over them is implied. So, the specific structure obtained by the mass matrix will depend of the form of the tensor $\varepsilon _{ij}$, which can be written as:

\begin{eqnarray}
\varepsilon _{ij}= \left|\varepsilon _{ij}\right|e^{i\phi _{ij}}.
\label{phase-tensor}
\end{eqnarray} 

As we will show below, the form that takes this parameter may generate hierarchical mass schemes, while appropriate mixing mass matrices can be obtained with some assumptions. By adopting the perturbation (\ref{alighment-perturb}), the mass factors in (\ref{eps2-4order}) become:

\begin{eqnarray}
\left(m_{\nu}^2(\epsilon ^2)\right)_{ij}&= &\frac{\upsilon ^2}{16}\left(\varepsilon _{(i)k}f_kf_k^{\ast}\varepsilon_{(j)k}^{\ast}\right)\epsilon^2f_if_j^{\ast} \nonumber \\
\left(m_{\nu}^2(\epsilon ^4)\right)_{ij}&= &- \frac{\upsilon ^2}{64}\left[\varepsilon _{(i)k}f_kf_k^{\ast}\varepsilon_{kl}^{\ast}f_l^{\ast}f_l+2\varepsilon _{(i)k}f_kf_k^{\ast}f_lf_l^{\ast}\varepsilon_{l(j)}^{\ast} \right. \nonumber \\
&&\left.+f_k^{\ast}
f_k\varepsilon_{kl}f_lf_l^{\ast}\varepsilon_{l(j)}^{\ast}\right]\epsilon^4f_if_j^{\ast}, \nonumber \\
\label{mass-expansion}
\end{eqnarray}
for each $ij$ component, and only a sum over the repeated indices $k$ and $l=1,2,3$ is assume. 

\section{Mass matrix structures}

\subsection{The basic bimaximal form}
In order to evaluate the predictability of the mechanism in the light of actual observations, we need to set out specific structures to the matrix $\varepsilon _{ij}$. For that, we explore a bimaximal form with NH scheme. First, in order to avoid unnatural small Yukawa couplings and fine-tuning, let us assume that all the three Yukawa couplings $f_i$ are of the order of unity. In that case, according to the hypothesis in (\ref{alighment-perturb}), the tensor structure of the $\Gamma$ couplings are governed completely by the tensor $\varepsilon _{ij}$. Second, as will be shown below, the exact bimaximal form in equation (\ref{bimax}) may be obtained if the off-diagonal components $ij=12$ and $13$ aligns exactly, i.e.:

\begin{eqnarray}
\varepsilon _{12}=\varepsilon _{13}=0.
\label{off-diag}
\end{eqnarray}
Also, the dominant term of the bimaximal matrix exhibits off-diagonal components with opposite sign in relation to the diagonal ones, feature that is inherited by the squared matrix. Thus, if we want to reproduce this feature, we would have to demand that $(m_{\nu}^2)_{22}=-(m_{\nu}^2)_{23}$. However, since this condition apply to the dominant terms, we demand it only to the $\epsilon ^2$ order of (\ref{mass-expansion}), i.e., 

\begin{eqnarray}
(m_{\nu}^2(\epsilon ^2))_{22}=-(m_{\nu}^2(\epsilon ^2))_{23}.
\label{sign-condition}
\end{eqnarray}
One way to achieve this, is making the $\varepsilon _{23}$ component to have a relative phase of $\pi$ in relation to the diagonal components $\varepsilon _{22}$ and $\varepsilon _{33}$:

\begin{eqnarray}
\phi _{22}-\phi _{23}= \phi _{23}-\phi _{33}=\pi,
\label{desfase}
\end{eqnarray}
so that after applying the condition (\ref{sign-condition}), we obtain the constraint:

\begin{eqnarray}
\left|\varepsilon _{22}\right|\left(\left|\varepsilon _{22}\right|-\left|\varepsilon _{23}\right|\right)=\left|\varepsilon _{23}\right|\left(\left|\varepsilon _{33}\right|-\left|\varepsilon _{23}\right|\right).
\label{bimax-constraint}
\end{eqnarray}

Finally, in addition to the exact alignment in (\ref{off-diag}), we align the component $11$ by making $\varepsilon _{11}=0$, while all the other components are perturbated in the same quantity, i.e.:

\begin{eqnarray}
\left|\varepsilon _{22}\right|=\left|\varepsilon _{23}\right|=\left|\varepsilon _{33}\right|=\left|\varepsilon _{0}\right|.
\label{bimax-alignment}
\end{eqnarray}
Thus, by replacing the above assumptions into (\ref{mass-expansion}), we obtain the basic bimaximal form in (\ref{bimax}) with $\alpha =\beta =0$, which is reproduced also by the squared mass matrix:

\begin{eqnarray}
m_{\nu }^{2(0)}=\frac{\upsilon ^2}{8}\epsilon ^2\left|\varepsilon _0\right|^2\begin{pmatrix}
0 & 0 & 0 \\
0 & 1 & -1 \\
0 & -1 & 1
\end{pmatrix},
\label{bimax-limit}
\end{eqnarray}
which only exhibits one eigenvalue different from zero, i.e., in this limit we reproduce the two massless neutrinos of the bimaximal form, and one active neutrino with mass:

\begin{eqnarray}
m_3^{2(0)}=\frac{\upsilon ^2}{4}\epsilon ^2\left|\varepsilon _0 \right|^2,
\label{mass-3}
\end{eqnarray}
while two mixing angles (defined in the standard parametrization shown in equation (\ref{PMNS})) are zero ($\theta _{12}^{(0)}=\theta _{13}^{(0)}=0$), and one is maximal ($\theta _{23}^{(0)}=\pi /4$). Thus, the bimaximal form can be seen as a projection spanned by the $\varepsilon _{ij}$ space with the specific alignments described by (\ref{off-diag}), (\ref{desfase}) and (\ref{bimax-alignment}).

\subsection{First order perturbation from bimaximal form}

In order to have better predictions with actual data from oscillation of neutrinos, we can slightly deviate from the above basic bimaximal form. One way to achive this is by producing an small unalignment of the $11$ component in (\ref{bimax-alignment}), and spliting the degenerated parameters in the same strenght; for example, let us consider an small increase of the $\varepsilon _{22}$ component so that:

\begin{eqnarray}
\varepsilon _{11}=\left|\varepsilon _{22}\right|-\left|\varepsilon _{23}\right|>0,
\label{unalignment}
\end{eqnarray}
and $\phi _{11}=\phi _{22}$. Due to the constraint from equation (\ref{bimax-constraint}), if we split $\left|\varepsilon _{22}\right|$ and $\left|\varepsilon _{23}\right|$, then $\left|\varepsilon _{33}\right|$ also splits as:

\begin{eqnarray}
\left|\varepsilon _{33}\right|-\left|\varepsilon _{23}\right|=\left(1+\frac{\varepsilon _{11}}{\left|\varepsilon _{23}\right|}\right)\varepsilon _{11}.
\label{unalignment-2}
\end{eqnarray}

For simplicity, we reparametrize the above perturbation by defining:

\begin{eqnarray}
\lambda = \frac{\varepsilon _{11}}{\left|\varepsilon _{23}\right|}, \ \ \ \  m_0^2=\frac{\upsilon ^2}{8}\left|\varepsilon _{23}\right|^2\epsilon ^2,
\end{eqnarray}
As a first approximation, we take up to order $\epsilon ^2$, so that the $m_{\nu }^2(\epsilon ^2)$ contribution of (\ref{mass-expansion}) leads us to the following mass matrix:

\begin{eqnarray}
m_{\nu }^{2(1)}=m_0^{2}
\begin{pmatrix}
\frac{1}{2}\lambda ^2 & 0 & 0 \\ \\
0 & p(\lambda )  & -p(\lambda )   \\ \\
0  & -p(\lambda )  & p(\lambda )\left[q(\lambda )-1\right]
\end{pmatrix},
\label{nearly-bimax}
\end{eqnarray}
which exhibits the eigenvectors and eigenvalues shown in equations (\ref{order-1-eigenvec}) and (\ref{1-order-masses}) of the appendix \ref{app:pert-theor}, from where we obtain the following squared mass differences:

\begin{eqnarray}
\Delta m_{21}^{2(1)}&=&\frac{1}{2}m_0^2\left[p(\lambda)\left(q(\lambda )-r(\lambda)\right)-\lambda ^2\right], \nonumber \\
\Delta m_{31}^{2(1)}&=&\frac{1}{2}m_0^2\left[p(\lambda)\left(q(\lambda )+r (\lambda)\right)-\lambda ^2\right], \nonumber \\
\Delta m_{32}^{2(1)}&=&m_0^2p(\lambda)r(\lambda),
\label{order-1-massdiff}
\end{eqnarray}
where $\Delta m_{ij}^{2(1)}=m_{i}^{2(1)}-m_{j}^{2(1)}$, and the functions are:

\begin{eqnarray}
p(\lambda )=1+\lambda+\frac{1}{2}\lambda ^2, \ \ \ \ \ q(\lambda )=2+\lambda ^2 \  \ \ \text{and} \ \ \ r(\lambda)=\sqrt{4+\lambda ^4},
\end{eqnarray} 
while the $23$ mixing angle corrects as:

\begin{eqnarray}
s_{23}^{(1)}&=&\frac{r(\lambda)-\lambda ^2}{\sqrt{2}\sqrt{4-\lambda ^2\left(r(\lambda )-\lambda ^2\right) }},
\label{1-order mixing}
\end{eqnarray}
with $s_{23}=\sin \theta$. However, we still obtain two zero mixing angles, so we must to consider the next correction, as discussed below.

\subsection{Second order perturbation from bimaximal form}{\label{sub:2-orderpermut}}

By considering the term $\epsilon ^4$ from (\ref{mass-expansion}), the mass matrix can be written as: 

\begin{eqnarray}
m_{\nu}^{2(2)}=m_{\nu}^{2(1)}+\epsilon ^2\delta m_{\nu }^2,
\label{2-order-mass matrix}
\end{eqnarray}
with $m_{\nu}^{2(1)}$ the same first-order mass matrix as (\ref{nearly-bimax}), and $\delta m_{\nu }^2$ the second order correction written in equation (\ref{2-order-pert}) in the appendix \ref{app:pert-theor}. As we show in detail in this appendix, we obtain the corrections to the neutrino mass differences and mixing angles obtained before, with $\delta m_{\nu }^2$ as the next order perturbation of $m_{\nu}^{2(1)}$, and $\epsilon ^2$ the small parameter that produces the perturbation. The corrections to the mass differences have the form

\begin{eqnarray}
\Delta m_{ij}^{2(2)}=\Delta m_{ij}^{2(1)}+\epsilon ^2 \delta _{ij},
\label{2-order-massdiff-correct}
\end{eqnarray}
where $\delta _{ij}$ is obtained in equation (\ref{2-order-diffmass-correction}) of the appendix \ref{app:pert-theor}, while the mixing angles are obtained from the perturbated eigenvectors $\psi _{i}^{(2)}$ shown in (\ref{2-order-eigen}) in the same appendix. By comparing with the rotation matrix in the standard parametrization in (\ref{PMNS}), we obtain the second-order mixing angles shown in (\ref{2-order-mixingangles}). More explicitly, the $13$ angle has the form:

\begin{table}
\centering
\begin{tabular}{lll} \hline
\ \ \ Parameter & \ \ \ 2$\sigma$ & \ \ \ 3$\sigma$ \\ \hline \\
$\Delta m_{21}^2$ [$10^{-5}$ eV$^2$] & $7.49^{+0.19}_{-0.17}$ & $7.37^{+0.59}_{-0.44}$ \\ \\
$\Delta m_{3\ell}^2$ [$10^{-3}$ eV$^2$] & $2.526^{+0.039}_{-0.037}$ & $2.56^{+0.13}_{-0.11}$ \\ \\
$\sin ^2\theta _{12}$ & $0.308^{+0.013}_{-0.012}$ & $0.297^{+0.057}_{-0.047}$ \\ \\
$\sin ^2\theta _{13}$& $0.02163^{+0.00074}_{-0.00074}$ & $0.0215^{+0.0025}_{-0.0025}$\\ \\
$\sin ^2\theta _{23}$ & $0.440^{+0.023}_{-0.019}$ & $0.425^{+0.19}_{-0.044}$ \\ \hline
\end{tabular} 
\caption{Neutrino oscillation parameters at 2$\sigma $ \cite{fits, nufit}, and $3\sigma $ \cite{PDG}.}
\label{tab:oscillation parameters} 
\end{table}

\begin{figure}
\centering
\includegraphics[scale=0.21]{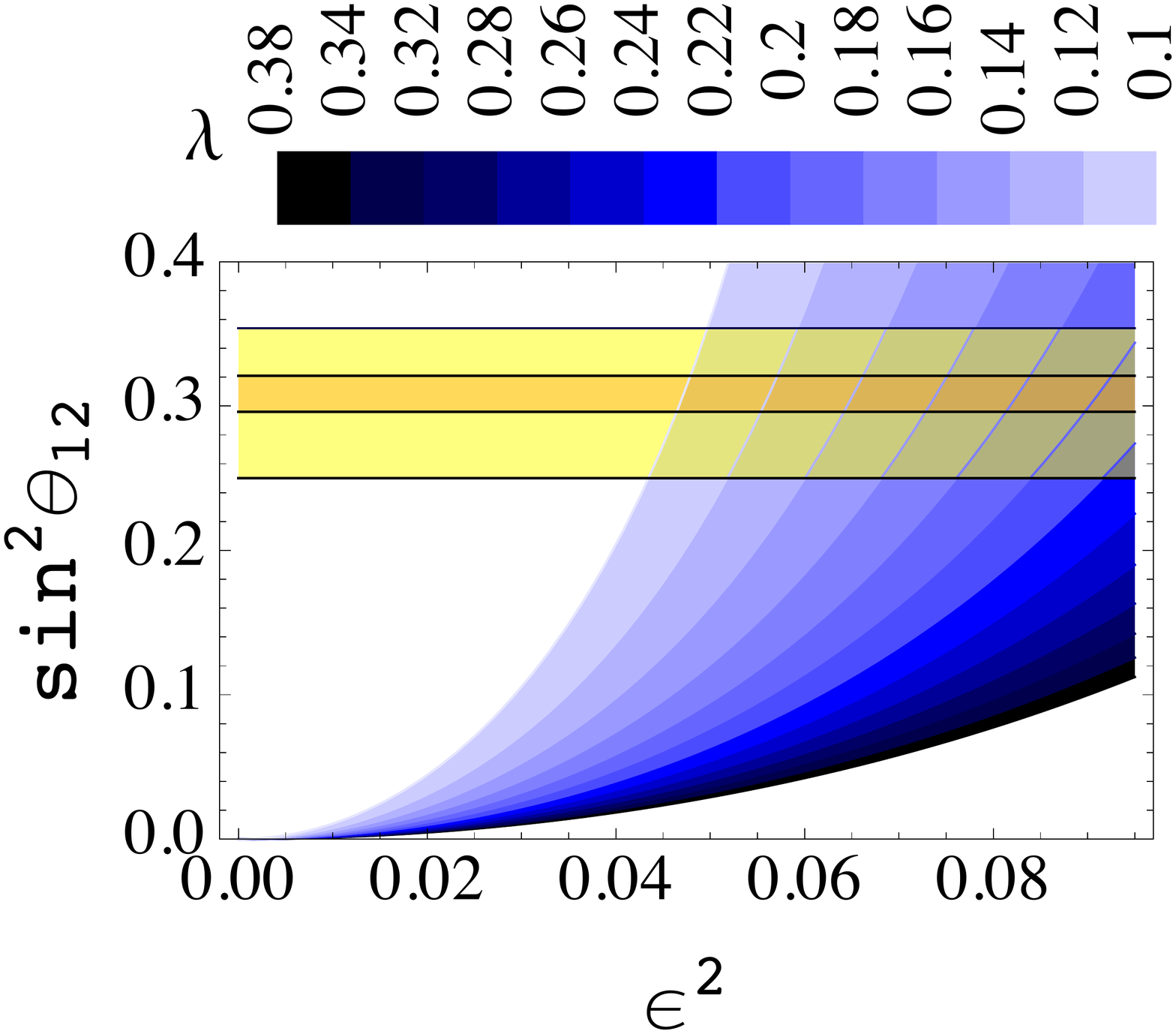}\hspace{-0.85cm}
\includegraphics[scale=0.21]{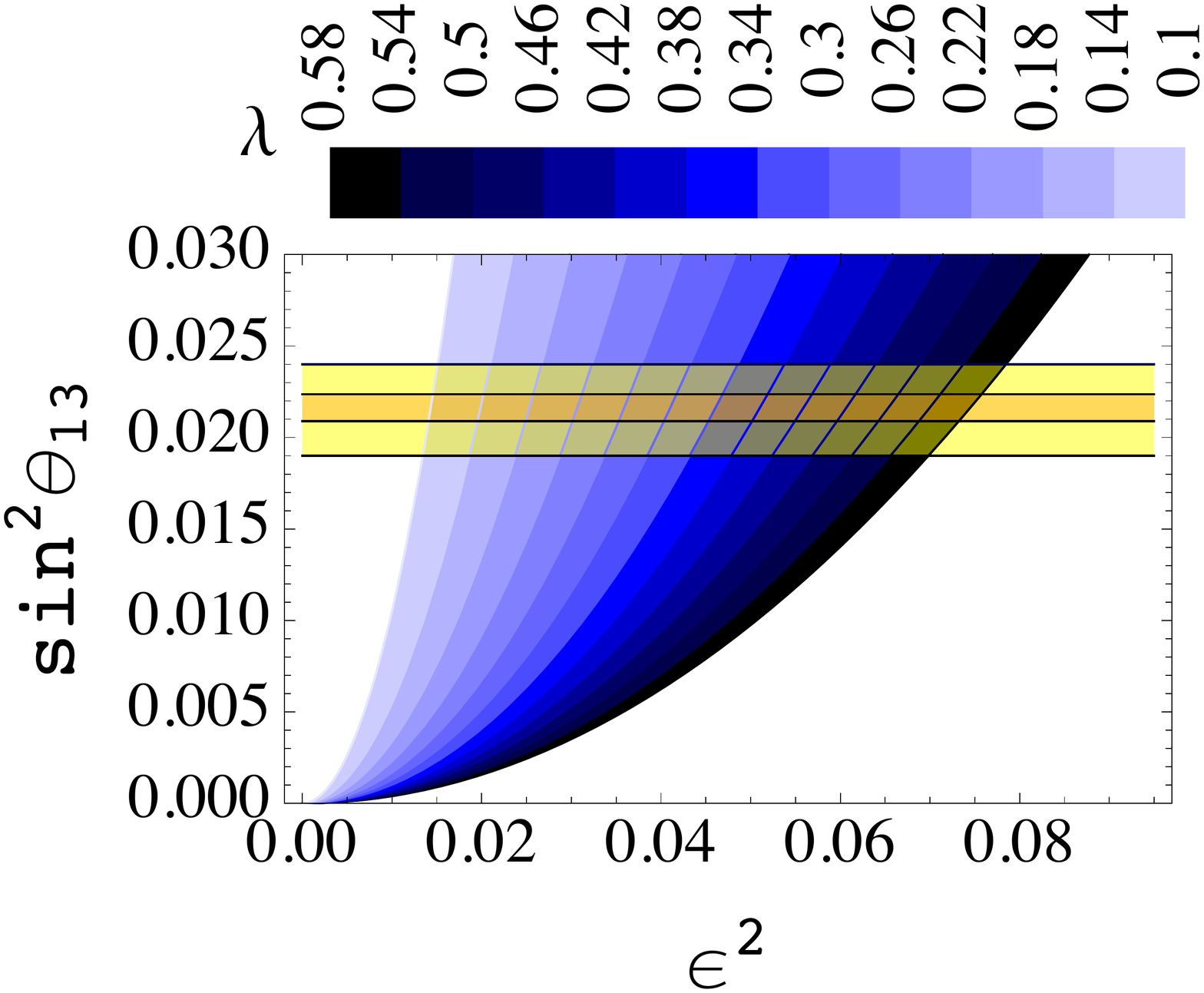}\hspace{-0.85cm}
\includegraphics[scale=0.21]{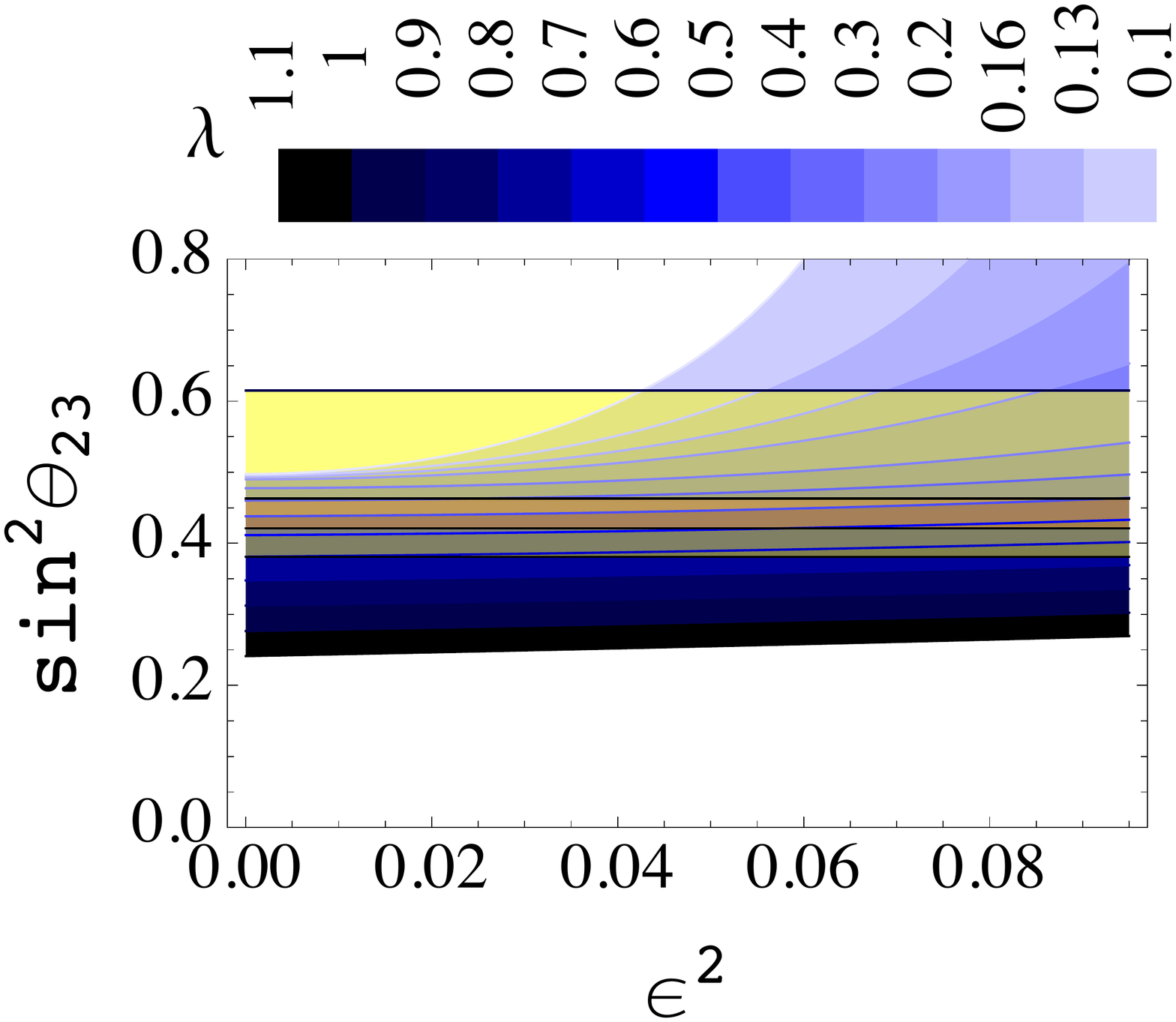}\vspace{-0.3cm}
\caption{Allowed region for $\epsilon = \upsilon /M_R$ and $\lambda =\varepsilon _{11} /\left|\varepsilon _{23}\right|$ compatible with the observed mixing angles at $2\sigma$ (red band ) and $3\sigma$ (yellow band) for NH scheme.}
\label{fig:eps-mix}
\end{figure}

\begin{eqnarray}
s_{13}^{(2)}=\epsilon ^2 \sum _{j\neq 1}\left(\frac{\psi_j^{(1)T}\delta m_{\nu}^2\psi_1^{(1)}}{\Delta m_{1j}^{2(1)}}\right)(\psi _j^{(1)})_{3},
\label{2-order-13angle}
\end{eqnarray}
while the $12$ and $23$ angles can be written as functions of the above as:

\begin{eqnarray}
s_{12}^{(2)}&=&\frac{\epsilon ^2}{\sqrt{1-(s_{13}^{(2)})^2}}\sum _{j\neq 1}\left(\frac{\psi_j^{(1)T}\delta m_{\nu}^2\psi_1^{(1)}}{\Delta m_{1j}^{2(1)}}\right)(\psi _j^{(1)})_{2}, \nonumber \\
s_{23}^{(2)}&=&\frac{1}{\sqrt{1-(s_{13}^{(2)})^2}}\left[s_{23}^{(1)}+\epsilon ^2\sum _{j\neq 2}\left(\frac{\psi_j^{(1)T}\delta m_{\nu}^2\psi_2^{(1)}}{\Delta m_{2j}^{2(1)}}\right)(\psi _j^{(1)})_{3}\right].
\label{2-order-2jangles}
\end{eqnarray}

In order to evaluate numerical fits of predictions (\ref{2-order-massdiff-correct}), (\ref{2-order-13angle}) and (\ref{2-order-2jangles}), we find allowed regions according to fittings from experimental observations reported in table \ref{tab:oscillation parameters} at $2\sigma$ and $3 \sigma$. The colored density curves in Figure \ref{fig:eps-mix} shows the predicted second-order function $\sin ^2 \theta ^{(2)}_{ij}$ as function of the parameter $\epsilon ^2$ and for different values of $\lambda$. The horizontal bands corresponds to the $2\sigma $ and $3\sigma$ experimental values. We observe that solutions are found for relatively large values of $\epsilon ^2$, at order between $10^{-2}$ and $10^{-1}$, which corresponds to ratios $\upsilon / M_R \sim 0.1 - 0.3 $, obtaining right-handed neutrinos with masses of the order of $0.8 - 2.5$ TeV. We also see that the ratio $\lambda = \varepsilon _{11}/\left|\varepsilon _{23}\right|$ has an important effect on the allowed regions. Although we do not impose any restriction on this parameter, values with $\lambda < 1$ favour smaller values of $\epsilon ^2$, which is important to preserve the perturbative nature of the analysis. In addition, the solutions are consistent with the bimaximal limit when $\lambda $ and $\epsilon $ approach zero, where $s^2 _{12}=s^2 _{13} = 0$, and $s^2 _{23}= 0.5$ (i.e., $\theta _{23}= \pi/4$). 

\begin{figure}
\centering
\includegraphics[scale=0.3]{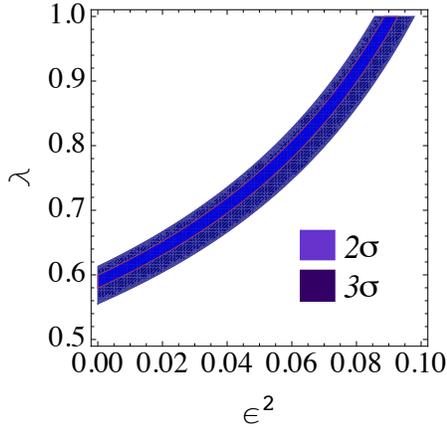}\vspace{-0.3cm}
\caption{Allowed region for $\epsilon = \upsilon /M_R$ and $\lambda $ compatible with the observed mass difference ratio $\Delta m_{21}^{2(2)}/\Delta m_{32}^{2(2)}$  at $2\sigma$ and $3\sigma$ for NH scheme.}
\label{fig:eps-lamb}
\end{figure}

Regarding the squared mass differences $\Delta m_{ij}^2$, we evaluate the ratio $\Delta m_{21}^{2(2)}/\Delta m_{32}^{2(2)}$, which is independent of $m_0$, and compare with the same ratio from the observed data. Plot in Figure \ref{fig:eps-lamb} shows the resulting allowed contourplot in the $\epsilon ^2 - \lambda$ plane. We see that $\epsilon ^2$ also ranges between $10^{-2}$ and $10^{-1}$, consistent with the regions obtained before for the mixing angles.

\section{Conclusions}

The see saw mechanisms help to understand the smallness of the neutrino masses by using a new large energy scale associated to a heavy particle, as for example, new scalar particles, sterile neutrinos or charged fermions. However, in most of the models either the scales of the new particle masses are restricted to very large values, unattainable for experimental verification in current particle accelerators, or their Yukawa couplings to the SM must fine-tune unnatural small values to obtain predictible parameters. 

In this paper, we have shown in a simple extension of the SM with massive neutrinos based in type I and II seesaw mechanism, that small active neutrino masses and their mixing angles can be reproduced according to observations, where the heavy neutrino is as light as the TeV scale while their Yukawa couplings to SM particles can be as large as unity. By enforcing an exact alignment condition between the Yukawa couplings, all the active neutrinos become massless, regardless the scale of the heavy neutrino. This cancellation remains at all orders of any perturbative expansion of the mass matrix unless the alignment breaks. Thus, perturbations of this condition is required to generate massive neutrinos, and the smallness of such perturbations become in the dominant mechanism responsible for small mass values. By adopting specific schemes as projections in the space spanned by the perturbation parameters, we can reproduce bimaximal forms for the neutrino masses. Deviations to the basic bimaximal structure was consider in order to obtain allowed regions of the parameters compatible with data from oscillation experiments.

\section*{Acknowledgment}

This work was supported by El patrimonio Aut\'onomo Fondo Nacional de Financiamiento para la Ciencia, la Tecnolog\'ia y la Innovaci\'on Fransisco Jos\'e de Calas programme of COLCIENCIAS in Colombia. F.O. thanks to the Theory Unit of the Physics Department of CERN, where part of this work was developed.

\appendix

\section{Block Diagonalization}\label{app:block}

Let us take a matrix of dimension four of the form:
\begin{eqnarray}
{\hat{M}}^2=
\begin{pmatrix}
A & B \\
B^{\dag} & C
\end{pmatrix},
\label{block-matrix}
\end{eqnarray}
with $A, B$ and $C$ sub-matrices of dimensions $3\times 3$, $1\times 3$ and $1\times 1 $, respectively, and $A\sim M_R^{-2}$, $B\sim M_R^{-1}$ and $C\sim 1$, so that they obey:
 
\begin{eqnarray}
A \ll B \ll C.
\label{block-hierarchy}
\end{eqnarray} 
The matrix (\ref{block-matrix}) can be block diagonalized by a unitary rotation of the form \cite{grimus}:

\begin{eqnarray}
V=
\begin{pmatrix}
\sqrt{1-FF^{\dag}} & F \\
-F^{\dag} & \sqrt{1-F^{\dag}F}
\end{pmatrix},
\label{approx-rotation}
\end{eqnarray}
where $F$ is a small sub-rotation with $F\ll 1$, and:

\begin{eqnarray}
V^T \hat{M}^2 V=\hat{m}^2=\begin{pmatrix}
a & 0 \\
0 & d
\end{pmatrix}
\label{diagonal-block}
\end{eqnarray}
a block diagonal matrix. In order to find the form of $F$, we must see the equation from the off-diagonal component of (\ref{diagonal-block}), which is:

\begin{eqnarray}
\sqrt{1-FF^{\dag}}AF+\sqrt{1-FF^{\dag}}B\sqrt{1-F^{\dag}F}-FB^{\dag}F-FC\sqrt{1-F^{\dag}F}=0.
\label{off-diagonal}
\end{eqnarray}
By expanding $F$ as inverse powers of $M_R$, $F=\sum _n F_n$, with $F_n\sim M_R^{-n}$, then we obtain the following expansion:

\begin{eqnarray}
\sqrt{1-FF^{\dag}}=1-\frac{1}{2}\sum _{n,m}F_nF_m^{\dag}-\frac{1}{8}\sum _{n,m,p,q}F_nF_m^{\dag}F_pF_q^{\dag} - ...
\label{root-expansion}
\end{eqnarray}
By replacing the above expansion into (\ref{off-diagonal}), we can obtain solutions for each order $F_n$. We do this up to fifth order. At first order, $M_R^{-1}$, equation (\ref{off-diagonal}) is $B-F_1C=0,$ obtaining:

\begin{eqnarray}
F_1=BC^{-1}.
\label{F1}
\end{eqnarray}
Up to order $M_R^{-2}$, the equation (\ref{off-diagonal}) become $B-F_1C-F_2C=0$, which by using the solution found in (\ref{F1}), we obtain that $F_2=0$. At order $M_R^{-3}$, and taking into account the above solutions, we obtain:

\begin{eqnarray}
F_3=ABC^{-2}-\frac{3}{2}BB^{\dag}BC^{-3}.
\label{F3}
\end{eqnarray}
As with $F_2$, the next even power, $F_4$, also cancels out. Finally, up to fifth order, equations (\ref{off-diagonal}), (\ref{F1}) and (\ref{F3}) leads us to the solution:

\begin{eqnarray}
F_5=A^2BC^{-3}-\left[\frac{5}{2}ABB^{\dag}+\frac{3}{2}BB^{\dag}\left(A+\frac{1}{3}A^{\dag}\right)\right]BC^{-4}+\frac{31}{8}BB^{\dag}BB^{\dag}C^{-5}.
\label{F5}
\end{eqnarray}

Regarding the diagonal components of (\ref{diagonal-block}), they give us the form of each block $a$ and $b$. In particular, we are interested in the lighter matrix $a$, which is:

\begin{eqnarray}
a=\sqrt{1-FF^{\dag}}A\sqrt{1-FF^{\dag}}-\sqrt{1-FF^{\dag}}BF^{\dag}-FB^{\dag}\sqrt{1-FF^{\dag}}+FCF^{\dag}.
\label{light-diagonal}
\end{eqnarray}
By using the expansion (\ref{root-expansion}), with the solutions for $F_{1,3,5}$ above, we obtain up to order $M_R^{-6}$ that:

\begin{eqnarray}
a&=&A-\left|B\right|^2C^{-1}-\frac{1}{2}\left(A\left|B\right|^2+\left|B\right|^2A\right)C^{-2}-\frac{1}{2}\left(A^2\left|B\right|^2+\left|B\right|^2A^{\dag}A-2\left|B\right|^4\right)C^{-3} \nonumber \\
&&+\frac{1}{2}\left(\left|B\right|^2A^{\dag}\left|B\right|^2-\left|B\right|^4A^{\dag}+\frac{7}{4}A\left|B\right|^4+\frac{11}{4}\left|B\right|^4A+\frac{3}{2}\left|B\right|^2A\left|B\right|^2 \right)C^{-4} \nonumber \\
&&-2\left|B\right|^6C^{-5}
\label{light-diagonal-2}
\end{eqnarray}
where $\left|B\right|^2=BB^{\dag}$. 

On the other hand, if the blocks in (\ref{block-matrix}) contains higher contributions, as in (\ref{mass-matrix-3}), the light mass matrix in (\ref{light-diagonal-2}) remains valid, but each matrix splits as $A=A_2+A_4$, $B=B_1+B_3$ and $C=C_0+C_2$, where the subindices label the power of $\epsilon \sim M_R^{-1}$. Thus, after separating each order, we found the following contributions:

\begin{eqnarray}
a(\epsilon ^2)&=&0,  \\
a(\epsilon ^4)&=&A_4-B_{13}C_0^{-1}+\left|B_1\right|^2C_2C_0^{-2}-\frac{1}{2}\{A_2,\left|B_1\right|^2\}C_0^{-2}+
\left|B_1\right|^4C_0^{-3} \label{light-mass-4} \\
a(\epsilon ^6)&=&-\left|B_3\right|^2C_0^{-1}+B_{13}C_2C_0^{-2}-\frac{1}{2}\{A_2,B_{13}\}C_0^{-2} -\frac{1}{2}\{A_4,\left|B_1\right|^2\}C_0^{-2} \nonumber \\
&&- \left|B_1\right|^2C_2^{2}C_0^{-3 }+\{B_{13},\left|B_1\right|^2\}C_0^{-3}+\{A_2,\left|B_1\right|^2\}C_2C_0^{-3} \nonumber \\
&&-\frac{1}{2}\{A_2^{2},\left|B_1\right|^2\}C_0^{-3}-3\left|B_1\right|^4C_2C_0^{-4}+\frac{7}{8}\{A_2,\left|B_1\right|^4\}C_0^{-4} \nonumber \\
&&+\frac{5}{4}\left|B_1\right|^2A_2\left|B_1\right|^2C_0^{-4}-2\left|B_1\right|^6C_0^{-5},\label{light-mass-6}
\end{eqnarray} 
with $B_{13}=B_1B_3^{\dag}+B_3B_1^{\dag}$, and \{\} denotes anticonmutator operations.

\section{Neutrino parameters}\label{app:neutrin-param}

For massive neutrinos, the flavour and mass eigenstates are related by the Pontercorvo-Maki-Nakagawa-Sakata mixing matrix, which in the standard parametrization is:

\begin{eqnarray}
V_{PMNS}=
\begin{pmatrix}
c_{12}c_{13} & s_{12}c_{13} & s_{13}e^{-i\delta} \\
-s_{12}c_{23}-c_{12}s_{23}s_{13}e^{i\delta } & c_{12}c_{23}-s_{12}s_{23}s_{13}e^{i\delta } & s_{23}c_{13} \\
s_{12}s_{23}-c_{12}c_{23}s_{13}e^{i\delta } & -c_{12}s_{23}-s_{12}c_{23}s_{13}e^{i\delta } & c_{23}c_{13}
\end{pmatrix},
\label{PMNS}
\end{eqnarray}
where $c_{ab}=\cos \theta _{ab}$ and $s_{ab}=\sin \theta _{ab}$. The mixing angle $\theta _{23}$ governs the oscillations of atmospheric neutrinos, $\theta _{12}$ the solar neutrinos, and $\theta _{13}$ that can be measured in short distance reactor neutrinos. The probabilities for oscillating also depends on the differences of the squared neutrino masses $\Delta m^2_{ij}=m^2_i-m^2_j$. Table \ref{tab:oscillation parameters} in subsection \ref{sub:2-orderpermut} summarize global fits at $2\sigma$ and $3\sigma$ from references \cite{fits} and \cite{PDG}, also available in NuFIT \cite{nufit}.

\section{Second order perturbation}\label{app:pert-theor}
  
Up to $\epsilon ^4$, the neutrino mass matrix has the form of equation (\ref{2-order-mass matrix}), where the second-order correction is:

\begin{eqnarray}
\delta m_{\nu}^2=-\frac{1}{2}m_0^2\lambda ^2
\begin{pmatrix}
1&\frac{1}{4}\left(1+\frac{2\lambda}{\left|\lambda \right|}\right) & \frac{1}{4}\left[3+\frac{2\lambda}{\left|\lambda \right|}+2\left(\lambda +\left|\lambda \right|\right) + \lambda ^2\right] \\ \\
\ast & \frac{1}{2} & \frac{1}{4}\left(4+4\lambda+\lambda ^2\right)\\ \\
\ast  & \ast  &  \frac{1}{2}\left(3+4\lambda+3\lambda ^2\right)
\end{pmatrix},
\label{2-order-pert}
\end{eqnarray} 
which can be seen as the correction to the ``unperturbated'' mass matrix $m_{\nu}^{2(1)}$ from (\ref{nearly-bimax}), so that the eigenvalues and eigenvectors correct as:

\begin{eqnarray}
m_i^{2(2)}&=&m_i^{2(1)}+\epsilon ^2\left(\psi_i^{(1)T}\delta m_{\nu}^2\psi_i^{(1)}\right), \nonumber \\
\psi_i^{(2)}&=&\psi_i^{(1)}+\epsilon ^2 \sum _{j\neq i}\left(\frac{\psi_j^{(1)T}\delta m_{\nu}^2\psi_i^{(1)}}{\Delta m_{ij}^{2(1)}}\right)\psi _j^{(1)},
\label{2-order-eigen}
\end{eqnarray}
with $m_i^{2(1)}$ and $\psi_i^{(1)}$ the eigenvalues and eigenvectors of the first-order matrix (\ref{nearly-bimax}), and $\Delta m_{ij}^{2(1)}=m_i^{2(1)}-m_j^{2(1)}$, where:

\begin{eqnarray}
\psi_1^{(1)}=\begin{pmatrix}
1 \\
0 \\
0
\end{pmatrix}, \ \ \ \
\psi_2^{(1)}=\begin{pmatrix}
0 \\
c_{23}^{(1)} \\
s_{23}^{(1)}
\end{pmatrix}, \ \ \ \
\psi_3^{(1)}=\begin{pmatrix}
0 \\
-s_{23}^{(1)} \\
c_{23}^{(1)}
\end{pmatrix},
\label{order-1-eigenvec}
\end{eqnarray}
with $s^{(1)}_{23}=\sin{\theta _{23}^{(1)}} $ defined in equation (\ref{1-order mixing}), and

\begin{eqnarray}
m_1^{2(1)}&=&\frac{1}{2}m_0^2\lambda ^2, \nonumber \\ 
m_2^{2(1)}&=&\frac{1}{2}m_0^2p(\lambda )\left(q(\lambda) - r(\lambda )\right), \nonumber  \\ 
m_3^{2(1)}&=&\frac{1}{2}m_0^2p(\lambda )\left(q(\lambda) +r(\lambda )\right).
\label{1-order-masses}
\end{eqnarray}

The second-order squared mass differences are:

\begin{eqnarray}
\Delta m_{ij}^{2(2)}=\Delta m_{ij}^{2(1)}+\delta _{ij},
\end{eqnarray}
with:

\begin{eqnarray}
\delta _{ij}=\epsilon ^2\left(\psi_i^{(1)T}\delta m_{\nu}^2\psi_i^{(1)}-\psi_j^{(1)T}\delta m_{\nu}^2\psi_j^{(1)}\right).
\label{2-order-diffmass-correction}
\end{eqnarray}

After putting all the above terms into the eigenvectors in (\ref{2-order-eigen}), we obtain the components of the neutrino rotation matrix, which have the form $V_{ij}=(\psi _i^{(2)})_j$. If we assume the standard parametrization as in (\ref{PMNS}) (by ignoring the CP phase), then we obtain the following second-order mixing angles:

\begin{eqnarray}
s_{13}^{(2)}&=&(\psi_1^{(2)})_3 \nonumber \\
s_{12}^{(2)}&=&\frac{(\psi_1^{(2)})_2}{\sqrt{1-(s_{13}^{(2)})^2}} \nonumber \\
s_{23}^{(2)}&=&\frac{(\psi_2^{(2)})_3}{\sqrt{1-(s_{13}^{(2)})^2}}.
\label{2-order-mixingangles}
\end{eqnarray}

\end{document}